\documentclass[a4paper,11pt]{article}
\usepackage{pos}
\usepackage{slashed}

\title{Proton Helicity GPDs from Lattice QCD}

\author*[a]{Joshua Miller}
\author[b]{Shohini Bhattacharya}
\author[c]{Krzysztof Cichy}
\author[a]{Martha Constantinou}
\author[d]{Xiang Gao}
\author[a]{Andreas Metz}
\author[e]{Swagato Mukherjee}
\author[e]{Peter Petreczky}
\author[f]{Fernanda Steffens}
\author[d]{Yong Zhao}

\affiliation[a]{Temple University, Philadelphia, PA 19122-1801, USA}
\affiliation[b]{RIKEN BNL Research Center, Brookhaven National Laboratory, Upton, NY 11973, USA}
\affiliation[c]{Faculty of Physics, Adam Mickiewicz University, ul. Uniwersytetu Poznańskiego 2, 61-614 Poznań, Poland}
\affiliation[d]{Physics Division, Argonne National Laboratory, Lemon, IL 60439, USA}
\affiliation[e]{Physics Department, Brookhaven National Laboratory, Upton, NY 11973, USA}
\affiliation[f]{Insitut für Strahlen- und Kernphysik, Rheinische Friedrich-Wilhelms-Universität Bonn, Nussallee 14-16, 53115 Bonn, Germany}

\emailAdd{joshua.miller0007@temple.edu}

\abstract{First lattice QCD calculations of $x$-dependent GPD have been performed in the (symmetric) Breit frame, where the momentum transfer is evenly divided between the initial and final hadron states. However, employing the asymmetric frame, we are able to obtain proton GPDs for multiple momentum transfers in a computationally efficient setup. In these proceedings, we focus on the helicity twist-2 GPD at zero skewness that gives access to the $\widetilde{H}$ GPD. We will cover the implementation of the asymmetric frame, its comparison to the Breit frame, and the dependence of the GPD on the squared four-momentum transfer, $-t$. The calculation is performed on an $N_f = 2+1+1$ ensemble of twisted mass fermions with a clover improvement. The mass of the pion for this ensemble is roughly 260 MeV.}

\FullConference{The 40th International Symposium on Lattice Field Theory (Lattice 2023)\\
July 31st - August 4th, 2023\\
Fermi National Accelerator Laboratory\\}

\begin{document}
\maketitle

\section{Introduction}
One of the most studied quantities for understanding the structure of strongly interacting particles is parton distribution functions (PDFs). Processes to measure PDFs include inclusive deep-inelastic lepton-nucleon scattering. The PDFs are one-dimensional objects, as they depend on the parton momentum fraction $x$, while the initial and final hadron states are identical. An extension of PDFs are the generalized parton distributions (GPDs), which also depend on $x$, but have an additional variable, the momentum transfer squared between the intial and final hadron states, $-t$. In addition, they depend on the so-called longitudinal momentum transfer, defined through the (skewness) parameter variable, $\xi$.

Accessing GPDs experimentally can be done through processes such as deeply virtual Compton scattering (DVCS)~\cite{DVCS,PhysRevD.55.7114}, deeply virtual meson production (DVMP)~\cite{DVMP}, as well as other processes with more complicated final states (such as in Ref.~\cite{Qiu_2022}). Processes such as DVCS and DVMP are extremely difficult to extract information from, and there has been a small number of experimental data sets compared to PDFs. Thus, conducting first principle calculations on the lattice is valuable. However, GPDs being defined on the light-cone make lattice QCD calculations difficult, except their first few Mellin moments. Novel methods, such as the quasi- and pseudo-distribution methods~\cite{quasi,lamet1,lamet2,pseudo}, use matrix elements for boosted hadrons with non-local operators, and then apply a matching formalism to relate to the light-cone distributions and their $x$-dependence.

One of the difficulties with lattice QCD is that calculations are computationally expensive and a large number of statistics is needed. Traditionally, calculations of GPDs in lattice QCD are performed in the symmetric (Breit) frame~\cite{unpol_hel,transversity,twist3}, which requires a new calculation for every value of the momentum transfer, $\vec{\Delta}$. Recently, a new Lorentz invariant parametrization of the GPDs and quasi-GPDs has been proposed in Ref.~\cite{unpol}, which is applicable in any frame. This enables a computationally efficient scheme of the calculations, since all the momentum transfer can be attributed to either initial or final state only, $\vec{p}_{i,f} = P_3\hat{z}-\vec{\Delta}$. This method allows us to cover a larger range of $-t$ within the same computational cost. 
In these proceedings, we apply this methodology for the helicity GPDs at zero skewness, obtaining $\widetilde{H}$ GPD. The complete calculation is presented in Ref.~\cite{Bhattacharya:2023jsc}. We employ the quasi-GPDs method, in which the Large Momentum Effective Theory (LaMET) is used to match the lattice data to the physical quantities. 

\section{Methodology}
GPDs are defined on the light-cone, and when studying the helicity case, we utilize the Dirac structure $\Gamma = \gamma^+\gamma_5$. On the light-cone, the axial matrix elements give two twist-2 GPDs, $\widetilde{H}$ and $\widetilde{E}$, are defined in position space according to Eq.~\eqref{eqn:ME_GPD}.
\begin{equation}
    F^{[\gamma^+\gamma_5]}(z^-,\Delta,P) = \bar{u}(p_f,\lambda')\bigg[\gamma^+\gamma_5\widetilde{H}(z^-,\xi,t) + \frac{\Delta^+\gamma_5}{2m}\widetilde{E}(z^-,\xi,t)\bigg]u(p_i,\lambda)\,.
    \label{eqn:ME_GPD}
\end{equation}
It should be noted that $\gamma^+$ is a linear combination of $\gamma^0$ and $\gamma^3$. When moving to the Euclidean definition of quasi-GPDs, we replace $+\rightarrow 3$, which is free from finite mixing under renormalization~\cite{Constantinou:2017sej}. In position space, we are finally left with Eq.~\eqref{eqn:ME_qGPD}.
\begin{equation}
    F^{[\gamma^3\gamma_5]}(z^3,\Delta,P) = \bar{u}(p_f,\lambda')\bigg[\gamma^3\gamma_5\widetilde{\mathcal{H}}_3(z^3,\xi,t;P^3) + \frac{\Delta^3\gamma_5}{2m}\widetilde{\mathcal{E}}_3(z^3,\xi,t;P^3)\bigg]u(p_i,\lambda)\,.
    \label{eqn:ME_qGPD}
\end{equation}
Looking at Eq. (\ref{eqn:ME_qGPD}), there are a few things to note. At zero-skewness, the right-hand side coefficient of $\widetilde{\mathcal{E}}_3$ is zero, and the latter drops out of the matrix element. 

In the new parametrization, the matrix elements are related to eight Lorentz invariant amplitudes, $\tilde{A}_i(z\cdot P, z\cdot \Delta, z^2, \Delta^2)$~\cite{Bhattacharya:2023jsc}, that is
\begin{align}
\widetilde{F}^{\mu} (z, P, \Delta)
& = \bar{u}(p_f,\lambda') \bigg [ \dfrac{i \epsilon^{\mu P z \Delta}}{m} \widetilde{A}_1 + \gamma^{\mu} \gamma_5 \widetilde{A}_2 + \gamma_5 \bigg ( \dfrac{P^\mu}{m} \widetilde{A}_3 + m z^\mu \widetilde{A}_4 + \dfrac{\Delta^\mu}{m} \widetilde{A}_5 \bigg ) \nonumber \\[0.1cm]
& \hspace{1.65cm} + m \slashed{z}\gamma_5 \bigg ( \dfrac{P^\mu}{m} \widetilde{A}_6 + m z^\mu \widetilde{A}_7 + \dfrac{\Delta^\mu}{m} \widetilde{A}_8 \bigg )\bigg ] u(p_i, \lambda) \, ,
\label{eqn:ME_Ai}
\end{align}
where $\epsilon^{\mu Pz\Delta} = \epsilon^{\mu\alpha\beta\gamma}P_{\alpha}z_{\beta}\Delta_{\gamma}$, $P^{\mu} = \frac{p_i + p_f}{2}\rightarrow P$, and $\Delta^{\mu}=p_f - p_i\rightarrow \Delta$. In order to obtain the matrix elements, we also consider the standard unpolarized and three polarized parity projectors; $\Gamma_{0} = \frac{1}{4}(1+\gamma_0)$ for unpolarized or $\Gamma_{\kappa} = \frac{i}{4}(1+\gamma_0)\gamma_5\gamma_{\kappa},~\kappa=1,2,3$. The complete expressions for the matrix elements for all projectors can be found in Ref.~\cite{Bhattacharya:2023jsc}. These are renormalized non-perturbatively in an RI-type scheme as described in the aforementioned paper. 
Using all the independent matrix elements, one can disentangle the amplitudes and obtain $\widetilde{\mathcal{H}}_3$ by comparing Eq.~\eqref{eqn:ME_Ai} and Eq.~\eqref{eqn:ME_qGPD}, leading to
\begin{equation}
\widetilde{\mathcal{H}}_3(\widetilde{A}_i;z) = \widetilde{A}_2 + P_3z\widetilde{A}_6 - m^2 z^2\widetilde{A}_7
    \label{eqn:quasi_H3}
\end{equation}
As discussed in  Ref.~\cite{Bhattacharya:2023jsc}, the definition of the quasi-GPDs, e.g., $\widetilde{\mathcal{H}}_3$, is not unique, which motivated the development of an alternative definition that is also Lorentz invariant, 
\begin{equation}
    \widetilde{\mathcal{H}} = \widetilde{A}_2 + P_3z\widetilde{A}_6\,.
    \label{eqn:quasi_H}
\end{equation}
In our kinematic setup, we obtain data for the $\pm$ directions of $P_3$, $\vec{\Delta}$ that lead to the same $-t$ value. We increase the statistics by using the symmetry properties of the Lorentz-invariant amplitudes under $z\rightarrow -z$, $P \to -P$, and $\Delta \to -\Delta$. We note that the matrix elements in the asymmetric frame do not have definite symmetry properties. 

One of the challenges in the calculation is related to the reconstruction of the $x$ dependence using a small number of discrete lattice data. To this end, we employ the Backus-Gilbert (BG) method ~\cite{BG} and we test the influence of $z_{max}$ in the final GPDs. The $x$ dependence reconstruction is followed by the matching formalism, in which we use the one-loop formula at zero skewness that is in a variant of RI ($\slashed{p}$)~\cite{OLM,match}.
\begin{equation}
    \label{eq:matching}
    q(x) = \int^{+\infty}_{-\infty} dy~f_1\left(\Gamma,y,\xi=0,\frac{p^z}{\mu}\right)_+ \tilde{q}(y)\,,
\end{equation}
\begin{equation}
    \label{eq:kernel}
f_1\left(\Gamma,y,\xi=0,\frac{p^z}{\mu}\right) = \frac{\alpha_sC_F}{2\pi} \left\{
        \begin{array}{ll}
            \frac{y^2+1}{x-1}\ln{\left(\frac{y}{y-1}\right)} - 1 & \quad  y< 0 \,, \\
            \frac{1+y^2}{1-y}\left[\ln{\frac{4y(1-y)(p^z)^2}{\mu^2}-1}\right]-2y+3 & \quad 0< y < 1 \,,\\
            -\frac{y^2+1}{y-1}\ln{\left(\frac{y}{y-1}\right)} + 1 & \quad y > 1\,.
        \end{array}
    \right.
\end{equation}
In Eq. (\ref{eq:matching}), $q(x)$ is the light-cone GPD, $\tilde{q}(y)$ is the quasi-GPD in momentum space, and\\
$f_1(\Gamma, y, \xi=0, \frac{p_z}{\mu})_+$ is the one-loop matching kernel as described by Eq. (\ref{eq:kernel}). 

\section{Lattice Calculations}

Our calculations are conducted on a lattice of dimensions $32^3 \times 64$ with a lattice spacing of $a=0.0934~{ \rm fm}$, corresponding to $L\approx 3~{ \rm fm}$. We use an $N_f = 2+1+1$ ensemble with twisted-mass fermions with a clover term and Iwasaki-improved gluons corresponding to $m_{\pi} = 260~{ \rm MeV}$. The ensemble we used has been generated by the Extended Twisted Mass Collaboration (ETMC)~\cite{AMG,twistedMass}.

The ingredients calculated on the lattice are two-point and three-point correlation functions. In order to produce results that have a good overlap with momentum-boosted proton states and suppress gauge noise, we use momentum smearing. In the symmetric frame, this is done for all different values of the momentum transfer. However, in the asymmetric frame where we group similar momentum transfers, e.g $(\Delta_x, 0, 0)$, we have chosen to optimize $\vec{\Delta}=(\pm2,0,0)$. The statistics obtained for the two-point and three-point correlation functions at each kinematic setup is shown in Table \ref{tab:stat}. The purpose of the symmetric frame calculation is to show that the numerical results respec the frame independence of the amplitudes, $\widetilde{A}_i$.

\begin{table}[h!]
\begin{center}
\renewcommand{\arraystretch}{1.2}
\begin{tabular}{lcccc|cccc}
\hline
frame & $P_3$ [GeV] & $\quad \mathbf{\Delta}$ $[\frac{2\pi}{L}]\quad$ & $-t$ [GeV$^2$] & $\quad \xi \quad $ & $N_{\rm ME}$ & $N_{\rm confs}$ & $N_{\rm src}$ & $N_{\rm tot}$\\
\hline
N/A       & $\pm$1.25 &(0,0,0)  &0   &0   &2   &329  &16  &10528 \\
\hline
symm      & $\pm$0.83 &($\pm$2,0,0), (0,$\pm$2,0)  &0.69   &0   &8   &67 &8  &4288 \\
symm      & $\pm$1.25 &($\pm$2,0,0), (0,$\pm$2,0)  &0.69   &0   &8   &249 &8  &15936 \\
symm      & $\pm$1.67 &($\pm$2,0,0), (0,$\pm$2,0)  &0.69   &0   &8   &294 &32  &75264 \\
symm      & $\pm$1.25 &$(\pm 2,\pm 2,0)$           &1.38   &0   &16   &224 &8  &28672 \\
symm      & $\pm$1.25 &($\pm$4,0,0), (0,$\pm$4,0)  &2.77   &0   &8   &329 &32  &84224 \\
\hline
asymm  & $\pm$1.25 &($\pm$1,0,0), (0,$\pm$1,0)  &0.17   &0   &8   &269 &8  &17216\\
asymm      & $\pm$1.25 &$(\pm 1,\pm 1,0)$       &0.34   &0   &16   &195 &8  &24960 \\
asymm  & $\pm$1.25 &($\pm$2,0,0), (0,$\pm$2,0)  &0.65   &0   &8   &269 &8  &17216\\
asymm      & $\pm$1.25 &($\pm$1,$\pm$2,0), ($\pm$2,$\pm$1,0) &0.81   &0   &16   &195 &8  &24960 \\
asymm  & $\pm$1.25 &($\pm$2,$\pm$2,0)          &1.24    &0   &16  &195 &8   &24960\\
asymm  & $\pm$1.25 &($\pm$3,0,0), (0,$\pm$3,0)  &1.38   &0   &8   &269 &8  &17216\\
asymm      & $\pm$1.25 &($\pm$1,$\pm$3,0), ($\pm$3,$\pm$1,0)  &1.52   &0   &16   &195 &8  &24960 \\
asymm  & $\pm$1.25 &($\pm$4,0,0), (0,$\pm$4,0)  &2.29   &0   &8   &269 &8  &17216\\
\hline
\end{tabular}
\caption{\small Statistics are shown for both frames with different momenta boosts and transfers, all with zero-skewness. The momentum unit $2\pi/L$ is 0.417 GeV. $N_{\rm ME}$, $N_{\rm confs}$, $N_{\rm src}$ and $N_{\rm total}$ are the number of matrix elements, configurations, source positions per configuration and total statistics, respectively.}
\label{tab:stat}
\end{center}
\end{table}

The three-point functions are calculated for operators containing a Wilson line in the direction of the momentum boost, $\hat{z}$. At each value of the length of the Wilson line, $z$, we utilize the ratio 
{\small{
\begin{equation}
\hspace*{-0.25cm}
R_\mu (\Gamma_\kappa, z, p_f, p_i; t_s, \tau) = \frac{C^{\rm 3pt}_\mu (\Gamma_\kappa, z, p_f, p_i; t_s, \tau)}{C^{\rm 2pt}(\Gamma_0, p_f;t_s)} \sqrt{\frac{C^{\rm 2pt}(\Gamma_0, p_i, t_s-\tau)C^{\rm 2pt}(\Gamma_0, p_f, \tau)C^{\rm 2pt}(\Gamma_0, p_f, t_s)}{C^{\rm 2pt}(\Gamma_0, p_f, t_s-\tau)C^{\rm 2pt}(\Gamma_0, p_i, \tau)C^{\rm 2pt}(\Gamma_0, p_i, t_s)}}
\label{eqn:Ratio}
 \end{equation}
}}
to extract the ground-state contribution of the matrix elements by a single-state fit. In this work, we use a source-sink time separation of $t_s = 10a$. 
In Eq.~\eqref{eqn:Ratio}, $\tau$ is the insertion time.

\section{Results}

\begin{figure}[h!]
    \centering
    \includegraphics[scale=0.35]{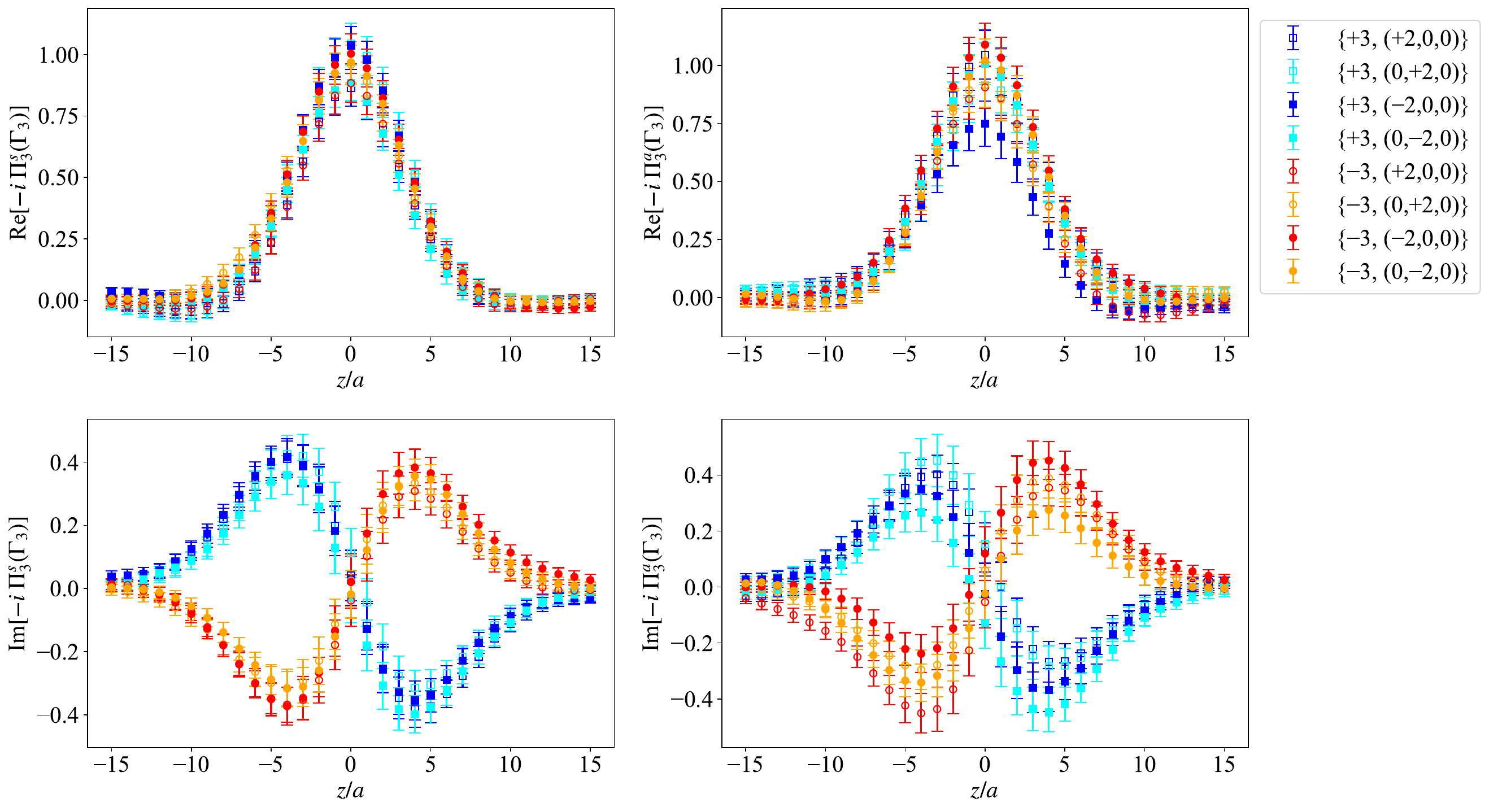}
    \vspace*{-0.75cm}
\caption{\small{Bare matrix element $\Pi_3(\Gamma_3)$ in the symmetric (left) and asymmetric frame (right), for $|P_3|=1.25$ GeV and $-t^s=0.69$ GeV$^2$ ($-t^a=0.65$ GeV$^2$) for the symmetric (asymmetric) frame. The top (bottom) panel corresponds to the real (imaginary) part. The notation in the legend is $\{P_3,\vec{\Delta}\}$ in units of $2\pi/L$.}}
    \label{fig:Pi3G3}
\end{figure}

We first present the bare matrix elements using Eq.~\eqref{eqn:Ratio} in the symmetric (``s'') and asymmetric frame (``a''). One of the dominant contributions in terms of signal is $\Pi_3^{s/a}(\Gamma_3)$, which corresponds to the operator $\gamma_3 \gamma_5$ using the polarized parity projector in the direction of the boost; this is shown in Fig. (\ref{fig:Pi3G3}) for $-t^s=0.69$ GeV$^2$ and $-t^a=0.65$ GeV$^2$ . As mentioned previously, the matrix elements calculated in the symmetric frame have definite symmetry properties, which does not hold in the asymmetric frame. Nevertheless, we find similarities between the asymmetric matrix elements in the eight kinematic setups leading to the same $-t$, implying that the asymmetries are small.
\begin{figure}[h!]
    \centering \includegraphics[scale=0.35]{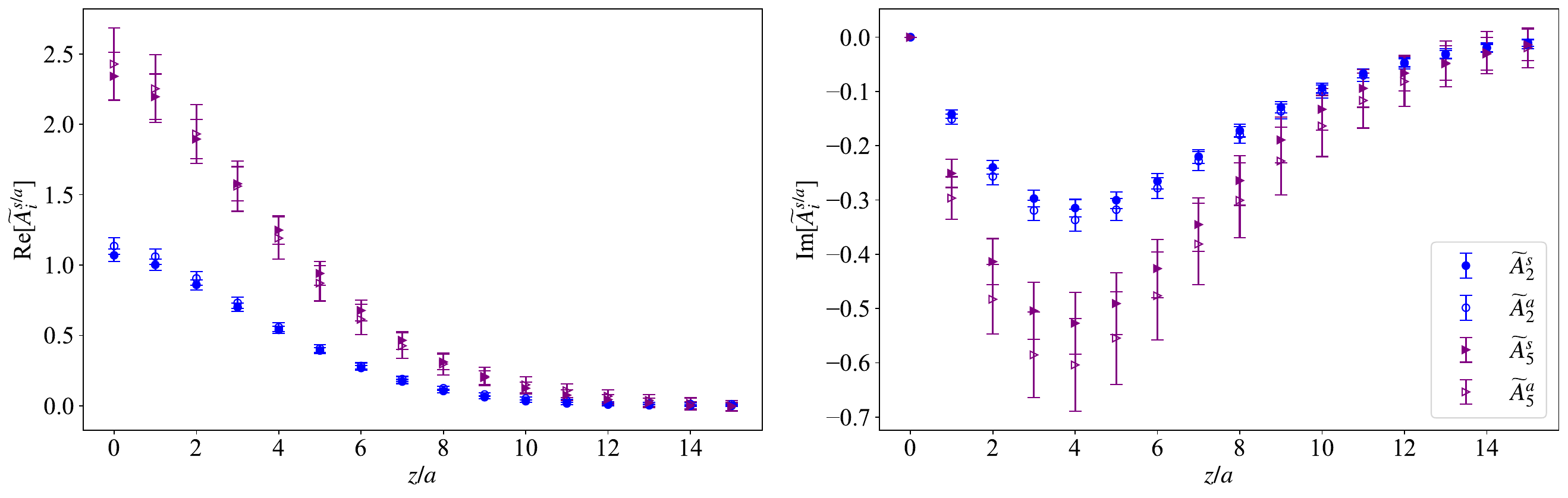}
    \vspace*{-0.75cm}
    \caption{\small Comparison of bare values of $\widetilde{A}_2$ and $\widetilde{A}_5$ in the symmetric (filled symbols) and asymmetric (open symbols) frame. The real (imaginary) part is shown in the left (right) column. The data correspond to $|P_3|=1.25$ GeV and $-t=0.69$ GeV$^2$ ($-t=0.65$ GeV$^2$) for the symmetric (asymmetric) frame. }
    \label{fig:Ai_a}
\end{figure}

After calculating the matrix elements, we are able to apply the decomposition into the amplitudes. For the same values of $-t$, we expect from theory that the amplitudes match. Here, we use periodic boundary conditions and, therefore, $t^a$ and $t^s$ cannot be matched exactly, but have $\sim 5\%$ difference for the values presented in Fig.~\ref{fig:Pi3G3}. However, this is very small and we anticipate that it does not have an impact on the comparison given the statistical uncertainties. In Fig. \ref{fig:Ai_a}, we compare the amplitudes with the largest signal, $\widetilde{A}_2$ and $\widetilde{A}_5$, in the two frames. We find that, within uncertainties, the values of the amplitudes do not depend on the frame that has been used in the calculation. The remaining amplitudes can be found in Ref.~\cite{Bhattacharya:2023jsc}.

\begin{figure}[h!]
    \centering
\includegraphics[scale=0.36]{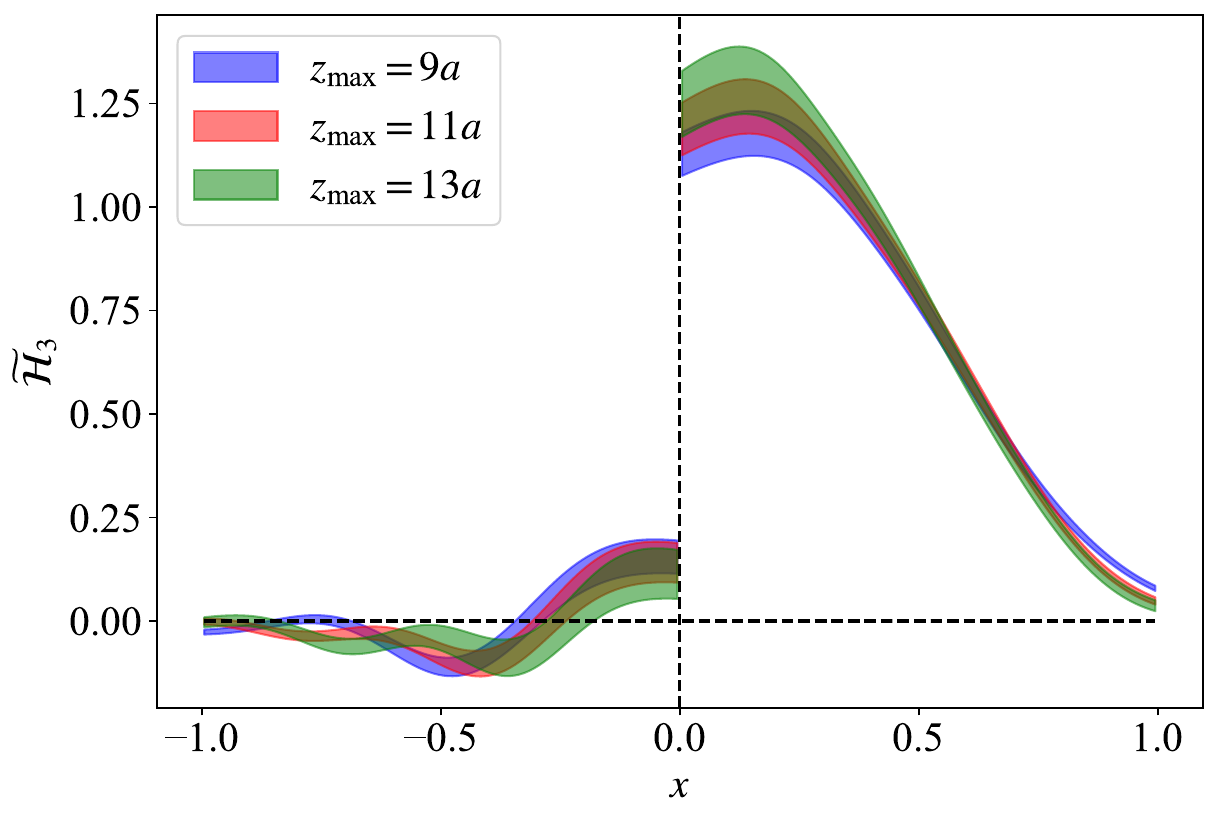}\quad
\includegraphics[scale=0.36]{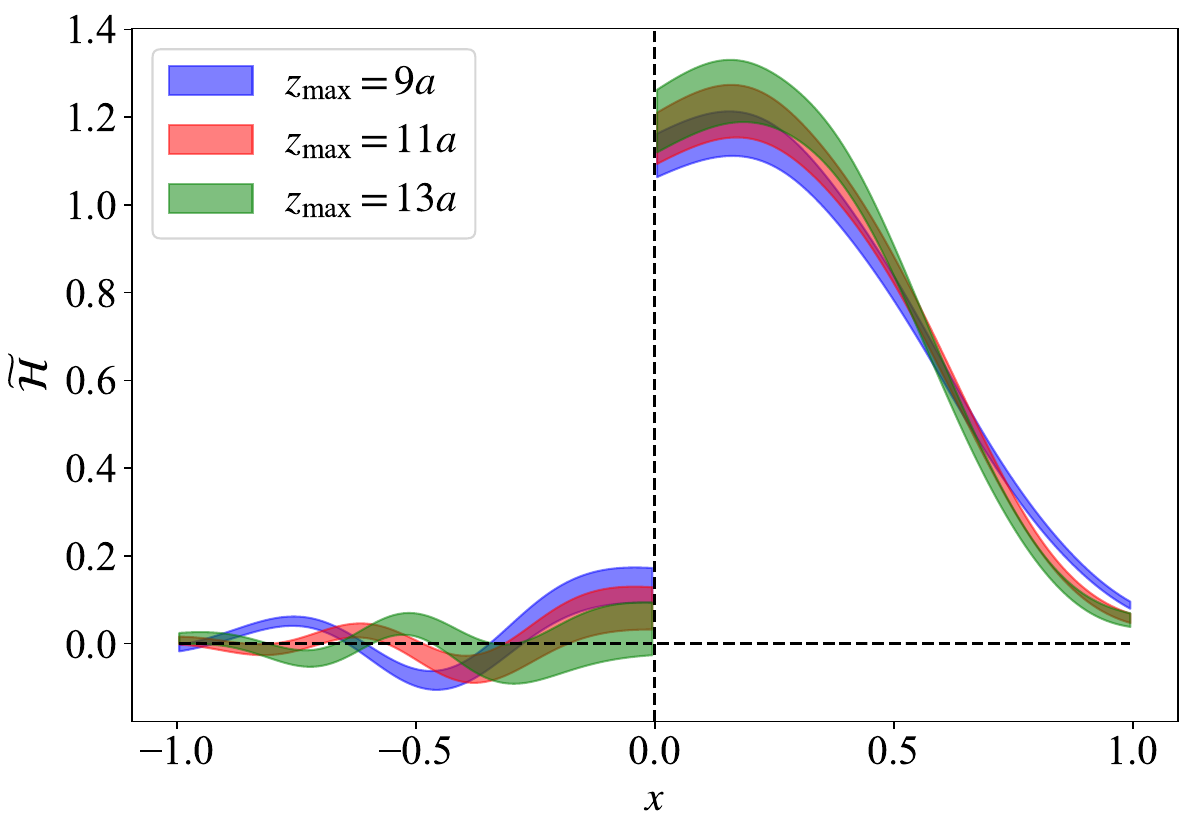}
    \vspace*{-0.55cm} 
    \caption{\small The $x$-dependent quasi-GPD $\widetilde{\cal H}_3$ (left) and $\widetilde{\cal H}$ (right) for $-t^a=0.65$ GeV$^2$ and $|P_3|=1.25$ GeV using Backus-Gilbert with $z_{\rm max}=9a,\,11a,\,13a$.}
    \label{fig:FH3_H_x_zmax}
\end{figure}

The next step of the work is the reconstruction of the $x$ dependence via a Fourier transform, or another reconstruction method. As mentioned previously, we apply the Backus-Gilbert reconstruction in order to minimize the impact of functional dependence. We test different values of $z_{max}$ as inputs into the reconstruction and look for stability. In Fig. \ref{fig:FH3_H_x_zmax}, we demonstrate our findings for both definitions of the $\widetilde{H}$-GPD. There are two kind of conclusions: (a) for each quasi-GPD, a consistency between all $z_{max}$ values of to $x=0.7$, and for $z_{max}=11a$ and $13a$ can be seen; (b) the two definitions are numerically very similar. In this work we use $z_{max}=11a$ as a final value. 
\begin{figure}[h!]
    \centering
\includegraphics[scale=0.29]{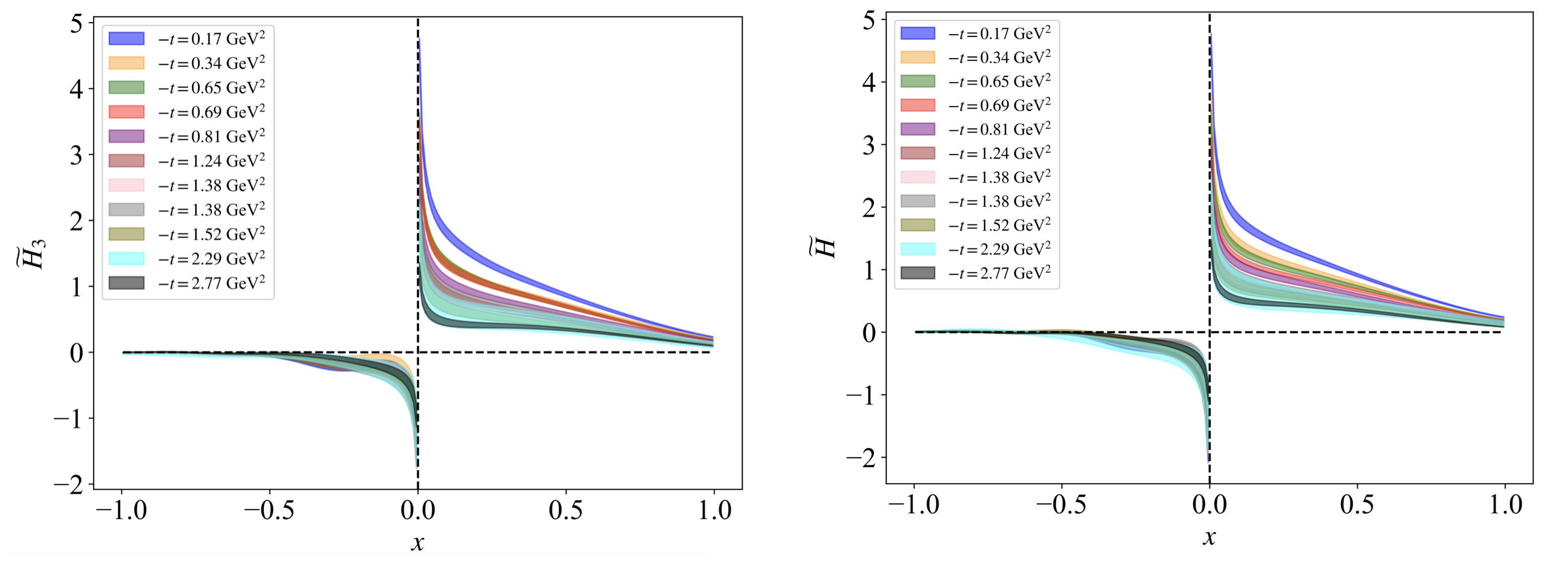}
    \vspace*{-0.75cm} 
    \caption{\small{The momentum-transfer squared dependence of the light-cone GPD $\widetilde{H}_3$ (left) and $\widetilde{H}$ (right) at $|P_3| = 1.25~{ \rm GeV}$. Results are given in the $\overline{{ \rm MS}}$ scheme at 2 GeV.}}
    \label{fig:t_dependence_GPD}
\end{figure}

Upon having the quasi-GPD in momentum space, we must now relate them to the light-cone GPDs using the LaMET matching formalism. We use the one-loop kernel at zero-skewness shown in Eqs.~\eqref{eq:matching} - \eqref{eq:kernel}. The matching is applied to all values of $-t$ presented in Table~\ref{tab:stat}, and the results are presented in Fig. \ref{fig:t_dependence_GPD}. In particular, we plot eleven different values of $-t \in [0.17,~2.77]~{ \rm GeV}^2$ at the $\overline{{ \rm MS}}$ scheme of 2 GeV. Both the standard definition, $\widetilde{H}_3$, and the alternative definition, $\widetilde{H}$, are shown. It should be noted that, given the value of the momentum boost, $|P_3|=1.25$ GeV, the GPDs extracted at $-t$ above $\sim 1$ GeV$^2$ suffer from enhanced higher-twist contaminations. Nevertheless, they are included in this study as they are obtained at no additional computational cost.

\section{Summary}
In these proceedings, we present a lattice calculation of the twist-2 axial vector $\widetilde{H}$ GPD for the proton at zero skewness; the $\widetilde{E}$ GPD is not accessible in this case due to a vanishing kinematic coefficient. Calculations presented here were performed on an $N_f=2+1+1$ ensemble with a clover term and twisted mass fermions. The pion mass for this ensemble is 260 MeV and the lattice spacing is $a=0.0934~{ \rm fm}$ and dimensions $32^3\times 64$. 
We applied the quasi-GPD approach as laid out in the Large-Momentum Effective theory (LaMET).
We explore two definitions of the quasi-GPDs, one which is the standard definition $\gamma_3\gamma_5$ and another Lorentz invariant version. We find that both definitions lead to numerically compatible results in the kinematic setups used in this work. The lattice data are renormalized using an RI-type prescription, and we employ the Backus-Gilbert approach to extract the $x-$dependence of the quasi-GPDs. For the matching to the light-cone GPDs, we use a one-loop matching from the RI to the $\overline{{ \rm MS}}$ at a scale of 2 GeV. In this analysis, we compared the numerical values of the Lorentz invariant amplitudes, $\widetilde{A}_i$ obtained in the symmetric and an asymmetric frame, where we demonstrated agreement for similar values of $-t$. Finally, we present the twist-2 axial vector GPD, $\widetilde{H}$ for eleven different values of $-t$. The results of this work can be used to parametrize the $-t$ dependence of the GPDs. The work can also be extended to nonzero skewness, as the framework of Ref.~\cite{Bhattacharya:2023jsc} supports any kinematic setup for the momentum transfer.

\section*{\centering Acknowledgements}
Tha Authors are grateful to Aurora Scapellato for her contributions to the initial stages of this project. S. B.,
S. M., and P. P. are supported by the U.S. Department of Energy, Office of Science, Office of Nuclear Physics
through Contract No. DE-SC0012704. S. B. is also supported by Laboratory Directed Research and Development
(LDRD) funds from Brookhaven Science Associates. K. C. is supported by the National Science Centre (Poland)
grants SONATA BIS no. 2016/22/E/ST2/00013 and OPUS no. 2021/43/B/ST2/00497. M. C., J. D., J. M. and A. S.
acknowledge financial support by the U.S. Department of Energy, Office of Nuclear Physics, Early Career Award
under Grant No. DE-SC0020405. J. D. also received support by the U.S. Department of Energy, Office of Science,
Office of Nuclear Physics, within the framework of the TMD Topical Collaboration. The work of A. M. has been
supported by the National Science Foundation under grant number PHY-2110472, and also by the U.S. Department
of Energy, Office of Science, Office of Nuclear Physics, within the framework of the TMD Topical Collaboration. F. S.
was funded by the NSFC and the Deutsche Forschungsgemeinschaft (DFG, German Research Foundation) through
the funds provided to the Sino-German Collaborative Research Center TRR110 “Symmetries and the Emergence
of Structure in QCD” (NSFC Grant No. 12070131001, DFG Project-ID 196253076 - TRR 110). Y. Z. and X. G.
received support by the U.S. Department of Energy, Office of Science, Office of Nuclear Physics through Contract
No. DE-AC02-06CH11357. X. G. was also partially supported by the U.S. Department of Energy, Office of Science,
Office of Nuclear Physics within the frameworks of Scientific Discovery through Advanced Computing (SciDAC)
award Fundamental Nuclear Physics at the Exascale and Beyond and the Quark-Gluon Tomography (QGT) Topical
Collaboration, under contract no. DE-SC0023646. Computations for this work were carried out in part on facilities
of the USQCD Collaboration, which are funded by the Office of Science of the U.S. Department of Energy. This
research used resources of the National Energy Research Scientific Computing Center, a DOE Office of Science
User Facility supported by the Office of Science of the U.S. Department of Energy under Contract No. DE-AC02-
05CH11231 using NERSC award NP-ERCAP0022961. This research was supported in part by PLGrid Infrastructure
(Prometheus supercomputer at AGH Cyfronet in Cracow). Computations were also partially performed at the Poznan
Supercomputing and Networking Center (Eagle supercomputer), the Interdisciplinary Centre for Mathematical and
Computational Modelling of the Warsaw University (Okeanos supercomputer), and at the Academic Computer Centre
in Gda´nsk (Tryton supercomputer). The gauge configurations have been generated by the Extended Twisted Mass
Collaboration on the KNL (A2) Partition of Marconi at CINECA, through the Prace project Pra13 3304 “SIMPHYS”.
Inversions were performed using the DD-$\alpha$AMG solver~\cite{AMG} with twisted mass support~\cite{twistedMass}.

\end{document}